\documentclass[superscriptaddress,prd,aps,preprint,eqsecnum,floats,showpacs]{revtex4}
\begin{document}
\title{The $\eta_c(3654)$ and  hyperfine splitting in charmonium}
\author{A.M. Badalian}
\affiliation{State Research Center,\\
 Institute of Theoretical and Experimental Physics, Moscow, Russia}
\author{B.L.G. Bakker}
\affiliation{Department of Physics and Astronomy, Vrije Universiteit,
 Amsterdam }
\def\la{\mathrel{\mathpalette\fun <}}
\def\ga{\mathrel{\mathpalette\fun >}}
\def\fun#1#2{\lower3.6pt\vbox{\baselineskip0pt\lineskip.9pt
\ialign{$\mathsurround=0pt#1\hfil ##\hfil$\crcr#2\crcr\sim\crcr}}}



\begin{abstract}
The hyperfine splitting for the 2S charmonium state is calculated and
the predicted number is $\Delta_{\rm HF}(2S) = 57 \pm 8$ MeV, being by
derivation the lower bound of this splitting. It results in
$M(\eta_c(2S))= 3630 \pm 8$ MeV, which is smaller by two standard
deviations than found in the Belle experiment \cite{ref.01}, but close
to the $\eta_c(2S)$ mass observed by the same group in the experiment
$e^+e^- \to J/\psi\, \eta_c$ \cite{ref.06} where $M(\eta_c(2S)) = 3622
\pm 12$ MeV was found.  
\end{abstract} 
\pacs{11.15.Bt, 13.25.Gv, 14.40.Gx}
\maketitle

\section{Introduction \label{sec.I}}
Recently the Belle Collaboration has    observed a new charmonium
state, the $\eta_c(2S)$, in exclusive $B\to KK_S K^-\pi^+$ decays
\cite{ref.01}. The measured mass of the $\eta_c(2S)$, $M(2^1S_0)=3654\pm 14$
MeV, has appeared to be rather close to the $\psi(2S)$ mass, $M_2=3686$
MeV, giving rise to a small hyperfine splitting for the $2S$ state,
\begin{equation}
\Delta_{\rm HF} (2S, \exp) =M_2-M(2^1S_0)=32\pm 14 \; {\rm MeV}.
\label{eq.01}
\end{equation} 
Compared to the hyperfine splitting for the $1S$ state,
$\Delta_{\rm HF} (1S, \exp) =117.2\pm 1.5$ MeV \cite{ref.02}, the
number (\ref{eq.01}) is 3.5 times smaller,
\begin{equation}
 \frac{\Delta_{\rm HF} (2S, \exp)}{\Delta_{\rm HF} (1S, \exp)}
 = 0.273\pm 0.123
\label{eq.02}
\end{equation} 
and has a large experimental error coming from the error in the
$\eta_c(2S)$ mass. However, in the ratio (\ref{eq.02}) even the upper
limit is rather small and it is of interest to compare this small
ratio with the predictions in the conventional theory of the spin-spin
interaction in QCD.

In this rapid communication we shall discuss the problem posed in a
model-independent way and show that  the experimental value
(\ref{eq.02}) puts a strong restriction on the coupling $\alpha_{\rm
HF}(2S)$, determining the spin-spin interaction. In particular, the
central value of $\Delta_{\rm HF}(2S)$ corresponds to $\alpha_{\rm
HF}(2S)\cong 0.18\div 0.19$ which implies the very large value for the
renormalization scale $\mu_2=2M_2\cong 7.4$ GeV. This scale appears to
be drastically different from that for the $1S$ state $(J/\psi)$ where
the scale in the strong coupling determining the HF interaction is
$\mu_1\cong \frac12 M_1=1.55$ GeV $(M_1=M(J/\psi))$ and  the coupling
constant $\alpha_{\rm HF} (\mu_1) \cong 0.335 $ is rather large. We
shall show that if one uses for the $2S$ state the same prescription as
for $J/\psi$, then our theoretical prediction is $\Delta_{\rm HF} (2S,
{\rm theory})=57 \pm 8$ MeV, which by derivation is the lower bound of
$\Delta_{\rm HF}(2S)$ and gives rise to the mass value $M(\eta_c(2S))= 3630 \pm
8$ MeV, being two standard deviations off the value obtained from the
Belle experiment~\cite{ref.01}.

\section{Model-independent calculation of $\Delta_{\rm HF} (2S)$ \label{sec.II}}
In one-loop approximation in the $\bar{MS}$ renormalization scheme,
the hyperfine splitting, $\Delta_{\rm HF}(nS)$, is given by the
well-known  expression:
\begin{equation}
\Delta_{\rm HF} (nS) =\frac89 \frac{\alpha_{\rm HF} (\mu_n)}{m^2_c}
|R_n(0)|^2\left(1+\frac{\alpha_{\rm HF}(\mu_n)}{\pi}
\xi_{\rm HF}\right),
\label{eq.03}
\end{equation} 
where the factor $\xi_{\rm HF}$ comes from the one-loop corrections, $
\xi_{\rm HF}=\frac{5}{12} \beta_0-\frac83-\frac34\ln 2$ \cite{ref.03},
and for $n_f=4$, $\xi_{\rm HF} (n_f=4)=0.2857$ is small, so that the
$\alpha_{\rm HF}$ correction in the brackets turns out to be less than
3\% $(\alpha_{\rm HF}(\mu_n)\la 0.35)$ and can be neglected in the
ratio (\ref{eq.02}).

For light mesons the relativistic version of the expression
(\ref{eq.03}) also exists, it was derived in Refs. \cite{ref.04} and
can be useful for charmonium, since  the current $c$-quark mass is not
large, $m_c=1.3.\pm 0.2$ GeV \cite{ref.02},
\begin{equation}
\Delta^R_{\rm HF} (nS) =\frac89 \frac{\alpha_{\rm HF}(\mu_n)}{\omega_n^2}
|R_n(0)|^2.
\label{eq.04}
\end{equation} 
Here $\omega_n$ is the kinetic energy matrix element:
\begin{equation}
\omega_n=\langle \sqrt{p^2+m^2_c}\rangle_{nS} ,
\label{eq.05}
\end{equation} 
which plays the role of the constituent quark mass. The HF splitting
(\ref{eq.03}) or (\ref{eq.04}) strongly depends on the $c$-quark mass
chosen and $\omega^2_n$ can change by a factor of 2 for different
choices of $m_c$.  To escape the problem of a correct choice of $m_c$
it is convenient to consider the ratio of hyperfine splittings with
relativistic corrections:
\begin{equation}
\frac{\Delta_{\rm HF}^R(2S)}{\Delta_{\rm HF}^R(1S)}
=\frac{\alpha_{\rm HF}(\mu_2)}{\alpha_{\rm HF} (\mu_1)}\left(
\frac{\omega_1}{\omega_2}\right)^2 \left|\frac{R_2(0)}{R_1(0)}
\right|^2.
\label{eq.06}
\end{equation} 
Note that the difference between $\omega_1$ and $\omega_2$ in
charmonium which takes into account relativistic corrections, is not
large, e.g., for the spinless Salpeter equation $(\omega_1 /
\omega_2)^2\cong 0.94$, however, it will be useful to keep this factor
in our analysis, since due to this factor the ratio (\ref{eq.06}) and
therefore the hyperfine splitting is 6\% smaller.  However, the
difference between $\omega(nS)$ and the current (pole) mass $m_c$
coming from relativistic corrections, is not small, being about 200 MeV
and 250 MeV for the 1S and 2S states, respectively (see
Section~\ref{sec.IV}).

The ratio of the wave functions at the origin occurring in
Eq.~(\ref{eq.06}) can be extracted in a model-independent way from the
leptonic widths:
\begin{equation}
 \Gamma_{e^+e^-}(nS) =\frac{4\alpha^2e^2_q}{M^2_n} |R_n(0)|^2\gamma_n ,
\label{eq.07}
\end{equation} 
where the factor $\gamma_n$ is 
\begin{equation}
 \gamma_n=1-\frac{16}{3\pi}
\alpha_s(M_n),
\label{eq.08}
\end{equation}  
and in $\gamma_n$ the strong coupling  $\alpha_s$  is taken at the
scale $\mu_n=M_n$ ($M_1=M(J/\psi)$, $M_2=M(\psi(2S))$. Note that just
this prescription for $\alpha_s(\mu)$ (in the QCD factor $\gamma$)
provides the minimal value of the extracted wave function at the
origin. For $\Lambda^{(4)}$(3-loop) $=280$ MeV ($n_f=4$), which
corresponds to $\alpha_s(M_z)=0.117$, the values of $\alpha_s(M_n)$ in
3-loop approximation are the following,
\begin{eqnarray}
 \alpha_s(M_1) & = & 0.247, \quad \alpha_s(M_2)=0.232, \nonumber \\
 \gamma_1 & = & 0.581, \quad \gamma_2=0.606. 
\label{eq.09}
\end{eqnarray}
Note that for another choice: $\gamma_1= \gamma_2$ which is often used,
the value of $\Delta_{\rm HF}(2S)$ would be 4\% larger than in our
consideration.

Then from the ratio of leptonic widths (\ref{eq.07}) it follows that
\begin{equation}
\left|
\frac{R_2(0)}{R_1(0)}\right|^2=\left(\frac{M_2}{M_1}\right)^2\frac{\Gamma_{e^+e^-}(2S)}{\Gamma_{e^+e^-}(1S)}
\frac{\gamma_1}{\gamma_2}. 
\label{eq.10}
\end{equation} 
Taking the experimental values: $\Gamma_{e^+e^-}(1S)=5.26\pm 0.37$ keV
and $\Gamma_{e^+e^-}(2S)=2.19\pm 0.15$ keV \cite{ref.02}, this ratio
with a good accuracy is
\begin{equation}
 \left|
\frac{R_2(0)}{R_1(0)}\right|^2=(0.59\pm 0.08)
\frac{\gamma_1}{\gamma_2}.
\label{eq.11}
\end{equation}  
The values of the wave functions at the origin extracted from the
leptonic widths are the following: $|R_1(0)|^2 = 0.917$ GeV${}^3$ and
$|R_2(0)|^2 = 0.518$ GeV${}^3$ and at this point it is important to
stress that for the $2S$ state the extracted value of the wave function
at the origin already takes into account the influence of the close
lying $D\bar{D}$ channel on the wave function in an implicit way, and
it is the same both for the hyperfine splitting and the leptonic width.
Due to the nearby $D\bar{D}$ threshold the experimental value of
$|R(0)|^2$ as well as the leptonic widths typically appear to be smaller
by about 20\% than in the theoretical calculations (one-channel
approximation) and the influence of the $D\bar{D}$ channel on the
$\psi(2S)$ mass was discussed in Ref.~\cite{ref.05}.

Now combining the expressions (\ref{eq.11}) and (\ref{eq.06}) one obtains
\begin{equation}
\frac{\Delta_{\rm HF}^R(2S)}{\Delta_{\rm HF}^R(1S)}=(0.59\pm0.08)\eta_{\rm HF}\label{eq.12}
\end{equation}
with the factor 
\begin{equation}
\eta_{\rm HF}=\left(\frac{\omega_1}{\omega_2}\right)^2\frac{\alpha_{\rm HF}(\mu_2)\gamma_1}{\alpha_{\rm HF}(\mu_1)\gamma_2}
.
\label{eq.13}
\end{equation}

From the analysis of the leptonic width and hyperfine splitting
for the $J/\psi$ it is known that the renormalization scale
$\mu_1$ in $\alpha_{\rm HF} (1S)$ is different from $M_1$ and a good
description of the  $\Delta_{\rm HF}(1S)$ can be reached if in 
Eq.~(\ref{eq.03}) or (\ref{eq.04}) the scale is taken to be $\mu_1\cong
\frac12 M_1=1.55$ GeV, which with $\Lambda^{(4)}(3-{\rm loop}) =280
$ MeV gives the value $\alpha_{\rm HF} (\frac12 M_1) =0.335$. Now with the
same prescription $\mu_2=\frac12 M_2$, the value $\alpha_{\rm HF} (\frac12
M_2)=0.307$  is obtained, and  then from Eqs.~(\ref{eq.09}) and
(\ref{eq.13}) it follows that
\begin{equation}
 \eta_{\rm HF} =0.826, \quad {\rm for}\quad
 \left (\frac{\omega_1}{\omega_2}\right)^2=0.94, \quad
 \frac{\gamma_1}{\gamma_2}  = 0.96.
\label{eq.14}
\end{equation}
As a result from Eq. (\ref{eq.12}) the "theoretical" ratio of the
hyperfine splittings is 
\begin{equation}
 \frac{\Delta_{\rm HF}^R(2S)}{\Delta_{\rm HF}^R(1S)}= (0.48 \pm 0.07)
\label{eq.15}
\end{equation} 
and our prediction for
the $\Delta_{\rm HF} (2S)$ is
\begin{equation}
\Delta_{\rm HF} (2S, {\rm theory}) =(57\pm 8) {\rm MeV}.
\label{eq.16}
\end{equation} 
Owing to our derivation this number can be considered as the lower
bound of $\Delta_{\rm HF}(2S)$ which was obtained in a
model-independent way.  The predicted mass of the $\eta_c(2S)$,
$M(\eta_c(2S)) = 3630 \pm 8$ MeV, differs by two standard deviations
from the measured value in the Belle experiment \cite{ref.01}, but is
rather close to the $\eta_c(2S)$ mass observed in another Belle
experiment \cite{ref.06} $e^+ e^- \to J/\psi\, \eta_c$ where the
measured $\eta_c(2S)$ mass, $M(\eta_c(2S))= 3622 \pm 12$ MeV is in good
agreement with our prediction.  Note that besides the dominant
perturbative term in the HF splitting there exists also a
nonperturbative (NP) contribution to $\Delta_{\rm HF} (nS)$. However,
the NP terms in charmonium can be calculated as in Ref.~\cite{ref.08}
and turn out to be small: we find $\Delta_{\rm HF}^{\rm NP}(1S)= 3-5$
MeV and $\Delta_{\rm HF}^{\rm NP}(2S)= 1-2$ MeV.  It is of interest to
note that the same model-independent estimate of the $\Delta_{\rm
HF}(2S)$ only with $\gamma_1= \gamma_2$ and $\alpha_{\rm HF}(\mu_1)=
\alpha_{\rm HF}(\mu_2)$ was suggested many years ago at the time when
the mass of the $\eta_c(1S)$ was not correctly measured \cite{ref.07}.

\section{The renormalization scale in $\alpha_{\rm HF}(\mu)$ \label{sec.III}}
To obtain the central value in Eq.~(\ref{eq.01}) in the same procedure
one needs to take the very small value $\alpha_{\rm HF} (\mu_2)=0.187$,
which corresponds to the large renormalization scale: $\mu_2=7.4 $ GeV
$\ga 2 M_2$. It is difficult to point out any physical explanation for such
different scales of $\mu_2=2 M_2$ for the 2S state and and $\mu_1\cong
\frac12 M_1=1.55$ GeV for the $J/ \psi$.

Existing theoretical calculations of $\Delta_{\rm HF}$ mostly provide
the $2S$ splittings in the range 70-90 MeV \cite{ref.09,ref.10}, while
only in Ref.~\cite{ref.11}, where the modified (screened) color Coulomb
interaction was used, the calculated HF splitting is smaller:
$\Delta_{\rm HF} (2S) =38$ MeV. However, in Ref.~\cite{ref.11}  with
the same modified  Coulomb interaction the fine structure splittings
for the $\chi_c(1P)$ mesons turned out to be twice as small as the
experimental data.

\section{Charmonium spectrum \label{sec.IV}}
Here we give as an illustration the spectrum and the constituent masses
$\omega_n$ in charmonium which were calculated solving the spinless
Salpeter equation with a linear plus Coulomb potential,
\begin{equation}
 \left( 2\sqrt{\hat p^2+m^2_c}+V_0(r)\right) \psi_{nL}(r)
 = M^{(0)}_{nL} \psi_{nL}(r). 
\label{eq.17}
\end{equation} 
In the static potential
\begin{equation}
 V_0(r) =-\frac43\frac{\alpha_{\rm st}}{r} +\sigma r
\label{eq.18}
\end{equation} 
the parameters $\alpha_{\rm st}=0.42$ and $\sigma=0.18$ GeV$^2$ were
taken.  For the one-loop pole mass we used $m_c=1.42$ GeV. The
remarkable feature of this set of parameters is that the meson mass, $
M{(nL)}= M^{(0)}{(nL)}-C_{SE}(nL) $,  contains a very small negative
subtractive constant which  can be strictly determined by the
nonperturbative self-energy contribution:  $C_{SE} = -4 \sigma\cdot
0.24 /\pi\omega_{nL}$ \cite{ref.12} and its value $C_{SE} =-33\pm 3$
MeV is small and approximately equal for all $nS$ states $(n\leq5)$.

\begin{table}
\caption{Spin-averaged masses $M(nL)$ for the spinless
Salpeter equation with the static potential $V_0(r)$ ($m_c=1.42$
GeV, $\sigma=0.18$ GeV, $\alpha_{\rm st}=0.42)$. \label{tab.01}}
\begin{tabular}{|l|l|l|l|}
\hline
 &$\omega_{nL} $(in GeV)& $M(nL)^{a)} $ in MeV& $M(nL)$ in MeV\\
 &&theory& experiment\\
\hline
 1S&1.64&3069&3067.6$\pm$0.05 \\
 2S&1.69&3662&3678.0$\pm$2.6\\
 3S&1.75&4083&4040$\pm$10\\
 4S&1.80&4433&4415$\pm$6\\
 1P&1.63&3526&3525.3$\pm$0.2\\
 2P&1.70&3967&---\\
 1D&1.66&3825&($1^3D_1$ state)\\
 &&&3769.9$\pm$2.5\\
 2D&1.72&4200& 4159$\pm$ 20?\\
\hline
\end{tabular}

$^{a)}$ The overall self-energy constant $C_{SE} =-33$ MeV is
used.\\

\end{table}

From Table~\ref{tab.01} one can see that the energy $\omega{(nS)}$ of
the $c$ quark, playing the role of the constituent mass  of a given
$nS$ state, appears to be around 1.65-1.75 GeV, i.e., essentially
larger than the current mass, $m_c\cong 1.4$ GeV in the Salpeter
equation  and gives rise to a small suppression of the HF splittings.
This fact can be considered as an explanation why in the
nonrelativistic approach the constituent $c$-quark mass is usually
taken to be larger, e.g., in Ref.~\cite{ref.09} $m_c=1.84$ GeV.  Note
also that with $\omega(1S)=1.60$ GeV and neglecting the NP contribution
of about 3-5 MeV, the value of $\alpha_{\rm HF}(1S)=0.355$ is needed in
order to obtain the experimental number for $\Delta_{\rm HF}(1S)=117$
MeV.

\section{Conclusions \label{sec.V}}
Thus we can conclude that the Belle experiment \cite{ref.01} on the
$\eta_c(2S)$ mass if it is confirmed, creates a number of interesting
theoretical problems, in particular, about the correct choice of the
renormalization scale of $\alpha_s(\mu)$ for the excited states. In the
conventional approach $\alpha_{\rm HF}(\mu_2) \cong 0.30$ and
$\Delta_{\rm HF}(2S)\cong (57 \pm 8)$ MeV and our prediction for the
mass is $M(\eta_c(2S))= 3630 \pm 88$ MeV which appears to be in
agreement with the result of the other Belle experiment \cite{ref.06}:
$e^+ e^- \to J/\psi,\eta_c$.  To obtain decisive conclusions it would
be important to measure the $\eta_c(2S)$ mass with better accuracy. The
confirmation of the value of $\Delta_{\rm HF}(2S)$ in the range close
to 32 MeV would require a drastic reconsideration of our understanding
of the scale in the HF interaction and possibly also of the QCD factor
$\gamma_n$ present in the leptonic width,  since they are
interconnected.

\begin{acknowledgments}
A.M.B. is grateful to L.B. Okun for his kind interest in this work
and wants to thank both him and A. Drutskoy for useful discussions.
This work was partly supported by the INTAS grant 00-00110 and
the RFFI grant 00-02-17836.
\end{acknowledgments}


\begin{thebibliography}{99}
\bibitem{ref.01}
S.-K. Choi et al., (the Belle Collaboration), Phys. Rev. Lett., {\bf
89}, 142001 (2002)

\bibitem{ref.02}
Particle Data Group, Phys. Rev., {\bf D66}, 010001-1  (2002)

\bibitem{ref.03} J. Pantaleone, S.-H.H Tye, and Y.Y. Ng, Phys. Rev., {\bf
D33}, 777 (1986)

\bibitem{ref.04} Yu.A. Simonov, Proc. XVII Int. Sch. Phys., Lisbon, 29
Sept. -- 4 Oct., 1999, p. 60, World Scientific (2000); hep-ph/9911239

\bibitem{ref.05} A. Martin ans J.M. Richard, Phys. Lett., {bf 115 B},
323 (1982)

\bibitem{ref.06} K. Abe et al., (the Belle Collaboration), hep-ex/0205104

\bibitem{ref.08} A.M. Badalian and B.L.G. Bakker,
Phys. Rev., {\bf D 64}, 114010 (2001);
hep-ph/0105156

\bibitem{ref.07} L.B. Okun and A.Yu. Khodzhamiryan, JETP Lett., {\bf
23}, 46 (1976)

\bibitem{ref.09} E.J. Eichten, K. Lane, and C. Quigg, Phys. Rev. Lett.,
{\bf 89} 162002 (2002)

\bibitem{ref.10}
G.S. Bali et al., Phys. Rev. {\bf D56}, 2566 (1997)\\ D. Ebert et al.,
Phys. Rev. {\bf D62}, 034014 (2000)\\ E.Y. Eichten and C. Quigg,
Phys. Rev. {\bf D49}, 5845 (1994)

\bibitem{ref.11} T.A. Lahde and D.O. Riska, Nucl. Phys. {\bf A707}, 425
(2002); hep-ph/0112131

\bibitem{ref.12} Yu.A. Simonov, Phys. Lett. {\bf B515}, 137 (2001);
hep-ph/0105141

\end{thebibliography}
\end{document}